\documentclass[11pt,twoside]{article}
\usepackage{asp2010}

\resetcounters

\begin{document}

\title{Revisiting the Impact of Axions in the Cooling of White Dwarfs}
\author{Brenda Melendez,$^{1,2}$ Marcelo Miller Bertolami,$^{1,2,3}$ and Leandro  Althaus$^{1,2}$
\affil{$^1$Facultad de Ciencias Astron\'omicas y Geof\'isicas, Universidad Nacional de La Plata, Paseo del Bosque s/n, 1900 La Plata, Argentina}
\affil{$^2$Instituto de Astrof\'isica de La Plata, UNLP-CONICET, Paseo del Bosque s/n, 1900 La Plata, Argentina}
\affil{$^3$Max-Planck-Institut f\"ur
  Astrophysik, Karl-Schwarzschild-Str. 1, 85748, Garching,
  Germany.}}

\begin{abstract}
It has been shown that the shape of the luminosity function of white
dwarfs can be a powerful tool to check for the possible existence of
DFSZ-axions. In particular, Isern et al. (2008) showed that, if the
axion mass is of the order of a few meV, then the white dwarf
luminosity function is sensitive enough to detect their existence. For
axion masses of about $m_a > 5$ meV the axion emission can be a
primary cooling mechanism for the white dwarf and the feedback of the
axion emission into the thermal structure of the white dwarf needs to
be considered.  Here we present computations of white dwarf cooling
sequences that take into account the effect of axion emission in a
self consistent way by means of full stellar evolution
computations. Then, we study and discuss the impact of the axion
emission in the white dwarf luminosity function.
\end{abstract}

\section{Introduction}

The Peccei-Quinn mechanism is one of the most convincing explanations
for the absence of CP-violating effects arising from the QCD vacuum
structure (see \citealt{1996slfp.book.....R}). A natural consequence
of this mechanism is the existence of a new particle, a boson called
the {\it axion} \citep{1978PhRvL..40..279W}. One of the axion models,
the DFSZ model \citep{Zhitnitsky, 1981PhLB..104..199D} allows for the
interaction of axions with charged leptons. The existence of such
DFSZ-axions would have an impact in the cooling of white dwarfs as
pointed by \cite{1986PhLB..166..402R} and
\cite{1992ApJ...392L..23I}. Because the evolution of white dwarfs is
mostly a simple cooling process and the basic physical ingredients
needed to predict their evolution are relatively well known, white
dwarfs offer a unique oportunity to test new physics under conditions
that can not be obtained in present day
laboratories. \cite{2008ApJ...682L.109I} showed that with the current
knowledge of the white dwarf luminosity function
\citep{2006AJ....131..571H, 2008AJ....135....1D} it might be possible
to detect axions as light as $m_a\sim 5$meV.  For axion masses which
might be detectable through the white dwarf luminosity function ($m_a>
5$meV) the axion cooling is comparable to the neutrino and photon
cooling of the white dwarf \citep{2008ApJ...682L.109I}.  In such
a situation, we expect the existence of a signifficant axion emission to
impact the thermal structure of the white dwarf, and consequently to
alter both the photon and neutrino emission. Thus, for the range of
interest of the axion masses we expect that the axion emission can not
be treated as a perturbation to the white dwarf cooling and that a
self consistent treatment of the axion emission is necessary. In the
present work, we improve previous works by studying the impact of the
axion emission in the cooling of white dwarfs by means of a self
consistent treatment of the axion emission and state of the art white
dwarf models.

\begin{figure}[]
\includegraphics[clip, angle=0, width=11.cm]{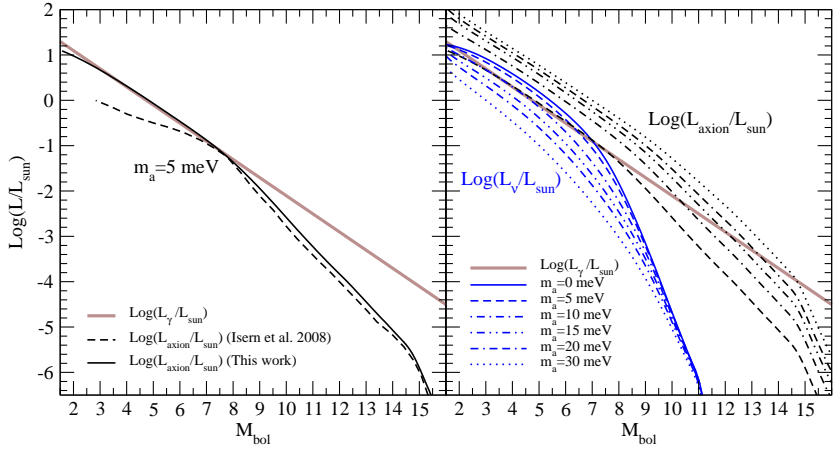} 
\caption{{\it Left:} Comparison of the axion ($m_a=5$meV) emission of
  our 0.609$M_\odot$ and the 0.61$M_\odot$ sequence of
  \cite{2008ApJ...682L.109I}. The effects of the departure from the
  isothermal core approximation can be apreciated at $M_{\rm bol}<6$.
  {\it Right:} Axion (black curves) and neutrino (blue curves) emission for our 0.609$M_\odot$
  sequences for different axion masses. The impact of the axion
  emission in the thermal structure of the white dwarf can be
  appreciated in the decrease of the neutrino emission at higher axion
  masses. Clearly, axion emission can not be treated perturbatively at
  $m_a>5$meV. }
\label{Fig:Lumi}
\end{figure}
\section{White dwarf models, input physics and the white dwarf luminosity function}
The initial white dwarf models were taken from
\cite{2010ApJ...717..183R}. Specifically 4 different initial white
dwarf models of 0.524$M_\odot$, 0.609$M_\odot$, 0.705$M_\odot$ and
0.877 $M_\odot$ were adopted. These models were obtained from
computing the complete evolution of initially 1$M_\odot$, 2$M_\odot$,
3$M_\odot$ and 5 $M_\odot$ ZAMS stars with Z=0.01, which is agreement
with semi-empirical determinations of the initial-final mass
relationship \citep{2009ApJ...692.1013S}. It is worth noting that
these sequences were computed from the ZAMS to the thermal pulses at
the AGB and finally to the post-AGB and white dwarf stages. For each
initial white dwarf model 6 cooling sequences with different assumed
axion masses were computed ($m_a$= 0, 5, 10, 15, 20 \& 30 meV). Axion
emission by both Compton and Bremsstrahlung processes was included,
although in white dwarfs only the latter is important. Axion
Bremsstrahlung emission under degenerate conditions was included
adopting the prescriptions of
\cite{1987ApJ...322..291N,1988ApJ...326..241N} for strongly coupled
plasma regime ($\Gamma >1$) and \cite{1995PhRvD..51.1495R} for weakly
coupled plasmas ($\Gamma <1$). Computations were performed with {\tt
  LPCODE} stellar evolution code, which is specifically tailored for
the computation of white dwarfs and includes all relevant
microphysics such as equation of state, radiative and conductive
opacities, element diffusion and can even handle the effects of phase
separation, crystallization and release of latent heat.

To construct theoretical white dwarf luminosity functions we followed
\cite{2008ApJ...682L.109I} but adopted the method presented by
\cite{1989ApJ...341..312I}. From the cooling times, $t_c(l,m)$,
computed with {\tt LPCODE} for each value of the axion mass, we
computed the number density of stars per luminosity as 
\begin{equation}
\frac{dn}{dl}=-\int_{M_1}^{M_2} \psi(t) \left(\frac{dN}{dM}\right)
 \left(\frac{\partial t_c}{\partial l}\right)_m dM
\end{equation}
For a given white dwarf luminosity bin ($l$), and mass of the
progenitor ($M$), the formation time of the star, $t$, is given by
$t+t_{\rm ev}(M)+t_c(l,m)=T_d$. To compute the white dwarf luminosity
function we adopt the following additional ingredients: A Salpeter
initial mass function $N(M)$, the initial-final mass relationship
$m(M)$ from \cite{2009ApJ...692.1013S}, the stellar lifetimes $t_{\rm
  ev}$ from the BaSTI database \citep{2004ApJ...612..168P} and
constant star formation rate $\psi$.  Normalization of theoretical
luminosity functions is done as in \cite{2008ApJ...682L.109I}.

\subsection{Results and future work}
\begin{figure}[]
\includegraphics[clip, angle=0, width=8.4cm]{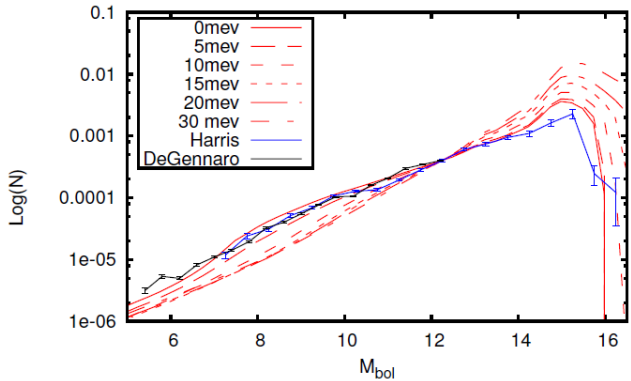} 
\caption{White Dwarf luminosity functions constructed for the
  different axion masses compared with the luminosity functions
  derived by \cite{2006AJ....131..571H} and
  \cite{2008AJ....135....1D}. DFSZ-axions heavier than $m_a>10$meV are
  clearly expcluded by the observed white dwarf luminosity functions.}
\label{Fig:WDLF}
\end{figure}
 
In Fig. \ref{Fig:Lumi} (left panel) the axion emission of our
0.609$M_\odot$ sequence is compared with the 0.61$M_\odot$ sequence of
\cite{2008ApJ...682L.109I} for an axion of $m_a=5$meV. There is an
overall good agreement between both predictions at low luminosities
($M_{\rm bol}>6$). The departure between both curves at high
luminosities ($M_{\rm bol}<6$) can be traced back to the isothermal
core approximation of \cite{2008ApJ...682L.109I} which leads to an
underestimation of the axion emission at high luminosities when the
maximum temperature of the core is located off-centered.  The right
panel of Fig. \ref{Fig:Lumi} shows both the axion and neutrino
emissions of our 0.609$M_\odot$ sequences computed with different
axion masses. As expected the higher the axion mass, the higher the
axion emission and the cooling speed. A more interesting feature can
be seen by analysing the neutrino emission at different axion
masses. As can be seen in Fig. \ref{Fig:Lumi} the neutrino emission is
reduced as the axion mass is increased. This result is due to the fact
that when axion emission is included this leads to an extra cooling of
the white dwarf core which alters the thermal structure of the white
dwarf (as compared with the case with no axion emission) which, in
turn, leads to a decrease of the neutrino emission at a given surface
luminosity of the star. Note that this is true even for our lighter
computed axion mass ($m_a=5$meV), for which the axion emission is
already different from the case with no axions ($m_a=0$meV). Then, the
inclusion of the feedback effects of the axion emission on the thermal
structure of the white dwarf, leads to a decrease of the neutrino
emission which will diminish the sensitivity of the white dwarf
cooling times to the axion mass.  Consequently the axion emission
needs to be treated self-consistently when dealing with axions in the
range detectable through the white dwarf luminosity function.  In
Fig. \ref{Fig:WDLF} we show the resulting white dwarf luminosity
functions for each axion mass as compared with the observationally
derived ones. It can be clearly seen that axion masses larger than 10
meV would lead to strong disagreements with the luminosity functions
derived by \cite{2006AJ....131..571H} and
\cite{2008AJ....135....1D}. On the other hand, the existence or not of
DFSZ-axions with masses lower than 5 meV can not be concluded without
a detailed statistical analysis of the uncertainties.

 We have shown that the impact of the axion emission on the white
 dwarf luminosity function will be overestimated if the effect of
 axion emission on the thermal structure of the white dwarf is not
 taken into account. This is particularly important at $\log
 L_\gamma>-1$.  Our preliminary analysis shows that the observed
 luminosity functions are consistent with the absence of any
 additional cooling mechanism (like axions), although a detailed
 statistical analysis of the results should be made before making a
 final statement. On the other hand, DFSZ-axion masses larger than
 $m_a> 10$ meV, are clearly excluded by our present knowledge of the
 white dwarf luminosity function, in agreement with
 \cite{2008ApJ...682L.109I}.


\acknowledgements M3B thanks the organizers of the EUROWD12 for the
finantial assistance that helped him to attend the conference.  This
research was supported by PIP 112-200801-00940 from CONICET and
PICT-2010-0861 from ANCyT.

\bibliographystyle{asp2010}
\bibliography{BM_M3B_LGA}
\end{document}